\documentclass[aps,prc,showpacs,superscriptaddress,preprint,byrevtex,floatfix]{revtex4-1}

\usepackage[dvips]{graphicx}
\usepackage{color}
\usepackage{amssymb}
\usepackage{amsmath}
\usepackage{braket}
\usepackage{hyperref}
\usepackage{bm}

\hypersetup{bookmarksnumbered,%
	colorlinks,%
	linkcolor=blue,%
	citecolor=blue,%
	plainpages=false,%
	pdfstartview=FitH}

\newcommand{\nc}{\newcommand}           
\nc{\vc}[1]     {\mbox{\boldmath $#1$}} 
\nc{\mapleft}[1]{                       
 \smash{\mathop{                        %
  \hbox to 0.90cm{\rightarrowfill} }\limits_{#1}}}
\nc{\figwidth}{0.55}                    

\nc{\mydraft}	{\setlength{\topmargin}{-1.5cm}}
\mydraft   
\begin{document}

\title{The tensor-optimized high-momentum antisymmetrized molecular 
dynamics with bare interaction and its application in ${}^{4}$He nucleus}

\author{Mengjiao Lyu}\email[]{mengjiao@rcnp.osaka-u.ac.jp}
\affiliation{Research Center for Nuclear Physics (RCNP), Osaka
University, Ibaraki, Osaka 567-0047, Japan}

\author{Takayuki Myo}\email[]{takayuki.myo@oit.ac.jp}
\affiliation{General Education, Faculty of Engineering, Osaka
Institute of Technology, Osaka, Osaka 535-8585, Japan}
\affiliation{Research Center for Nuclear Physics (RCNP), Osaka
University, Ibaraki, Osaka 567-0047, Japan}

\author{Masahiro Isaka}
\affiliation{Hosei University, 2-17-1 Fujimi, 
Chiyoda-ku, Tokyo 102-8160, Japan}

\author{Hiroshi Toki}
\affiliation{Research Center for Nuclear Physics (RCNP), Osaka
University, Ibaraki, Osaka 567-0047, Japan}

\author{Kiyomi Ikeda}
\affiliation{RIKEN Nishina Center, Wako, Saitama 351-0198, Japan}

\author{Hisashi Horiuchi}
\affiliation{Research Center for Nuclear Physics (RCNP), Osaka
University, Ibaraki, Osaka 567-0047, Japan}

\author{Tadahiro Suhara}
\affiliation{Matsue College of Technology, Matsue 690-8518, Japan}

\author{Taiichi Yamada}
\affiliation{Laboratory of Physics, Kanto Gakuin University,
Yokohama 236-8501, Japan}
\date{\today}

\begin{abstract}
We formulate the ``tensor-optimized high-momentum antisymmetrized molecular
dynamics (TO-HMAMD)'' framework for {\it ab initio} calculations of nuclei by
hybridizing the tensor-optimized (TO-) and high-momentum (HM-) AMD approaches.
This hybrid approach has advantages in both analytical simplicity and numerical
efficiency comparing with other AMD-based methods which treat the bare
interaction, especially for heavier nuclear systems. In this work, the $s$-shell
nucleus $^{4}$He is calculated with TO-HMAMD by including up to double product
of nucleon-nucleon ($NN$) correlations, described by using high-momentum pairs
and spatial correlation functions of nucleons. The total energy and radius of the 
$^{4}$He nucleus are well reproduced using the AV8$^\prime$ interaction. The
spin-isospin channel dependence is also discussed for $NN$-correlations, which
are found to be mostly contributed in the even-state channels, especially the
triplet-even channel. Analyses of the analytical formation and numerical results
suggest that TO-HMAMD could be a promising framework for $p$-shell nuclear
systems.
\end{abstract}


\maketitle

\section{Introduction} 
The bare nucleon-nucleon ($NN$) interaction has been determined
phenomenologically in high precision by reproducing the $NN$ scattering data
\cite{pieper01}. In recent years, it is in active progress to predict the $NN$
interaction from the underlying Quantum Chromodynamics (QCD) \cite{ishii07}. The
strong tensor force and short-range repulsion in the bare $NN$ interaction have
been observed in both phenomenological models and the underlying field theory.
In nuclei, strong $NN$-correlations, including tensor and short-range
correlations, are induced by tensor force and short-range repulsion,
respectively \cite{pieper01b}. In {\it ab initio} calculations, an accurate
description of $NN$ correlations is crucial for both the exact solution of
nuclear wave functions and the examinations of the underlying QCD predictions
for nucleon systems. Usually, in {\it ab initio} calculations, correlation
functions based on the Jastrow type or unitary transformation with the
exponential form are multiplied to the reference nuclear state \cite{pieper01,
neff03}.

In the present and previous works \cite{lyu18}, we propose the
``tensor-optimized high-momentum antisymmetrized molecular dynamics'' (TO-HMAMD)
for {\it ab initio} calculations of nuclear system, which is a hybridization of
the ``tensor-optimized antisymmetrized molecular dynamics" (TOAMD) method
\cite{myo15, myo17a, myo17b, myo17c, myo17d} and the ``high-momentum
antisymmetrized molecular dynamics" (HMAMD) method \cite{myo17e,myo18,isaka18}.
These three methods are based on the framework of ``antisymmetrized molecular
dynamics" (AMD), which has been very successful in microscopic description of
light nuclei, especially for the cluster states \cite{kanada03,kanada12}. In the
TOAMD approach, the variational correlation functions in operator forms are
explicitly formulated for both central and tensor channels and then their
multiple products successively act on the AMD basis state. The TOAMD method is
applied in {\it ab initio} calculations of $s$-shell nuclei by using the
AV8$^\prime$ bare interaction, and reproduces well the total energies and radii
of ${}^{3}$H and ${}^{4}$He nuclei \cite{myo17a}. In Ref.~\cite{myo17d}, it is
found that using the AV6 interaction, the TOAMD wave function provides better
energies for $s$-shell nuclei comparing with the variational wave functions with
the Jastrow type $NN$-correlations. In the HMAMD approach, the $NN$ correlations
are directly described by using the $NN$-pairs with large momenta in their
relative motion \cite{myo17e,myo18}, which are called "high-momentum pairs".
With these high-momentum pairs both of the tensor and short-range correlations
can be described satisfactorily. This approach is similar to the one used in
$i$SMT \cite{itagaki18} and AQCM-T \cite{matsuno18}, in which the tensor
correlation is treated. In Ref.~\cite{myo18}, energies and radii of the
${}^{3}$H nucleus are calculated through the HMAMD approach using multi
high-momentum pairs and the AV4$^\prime$ central interaction having short-range
repulsions. It is found that the solutions obtained in TOAMD and HMAMD
approaches converge exactly with each other and nicely reproduce the Green's
function Monte Carlo results \cite{myo18}. 

From the convergence behavior of the solutions for light nuclei between TOAMD
and HMAMD, it is concluded that the correlation function in the TOAMD method and
the high-momentum $NN$ pairs in the HMAMD method essentially describe the same
$NN$ correlations \cite{myo18}. However, there are differences of both physical
concepts and mathematical expressions between these two methods. If these two
methods are combined, we expect to get a better description of many-body
correlations in {\it ab initio} studies of nuclei, where many kinds of $NN$
correlations can be included in the wave function. We try to integrate the
analytical simplicity of the HMAMD approach and numerical efficiency of the
TOAMD approach. Hence, the hybridized nature of TO-HMAMD could be a promising
framework for $p$-shell nuclear system. In our previous work, we have applied
the TO-HMAMD method to the {\it ab initio} calculation of the simple ${}^{3}$H
nucleus, and discussed the accuracy and flexibility of the method. In this work,
we perform the detailed comparison of the formulations of TO-HMAMD with those of
HMAMD and TOAMD, and apply the TO-HMAMD method to {\it ab initio} calculations
of much more complicated nuclei ${}^{4}$He using the AV8$^\prime$ bare
interaction.

This paper is organized as follows. In Sec.~\ref{sec:formulation}, we introduce
the cluster expansion of the multiple products of the $NN$-correlations. We also
explain the HMAMD and TOAMD methods and finally the formulation of the
hybridized TO-HMAMD approach. We discuss in detail the advantages of the
TO-HMAMD method in comparison with HMAMD and TOAMD. In Sec.~\ref{sec:s-shell},
we present the numerical results and discussions of the {\it ab initio}
calculations for the $^{4}$He nucleus by using the TO-HMAMD method. In
Sec.~\ref{sec:corr}, we discuss the spin-isospin channel dependence of
$NN$-correlations in the wave function of TO-HMAMD. The last
Sec.~\ref{sec:conclusion} contains the conclusion.

\section{Formulation}
\label{sec:formulation}

We introduce the framework of TO-HMAMD by explaining first the AMD wave function
and next the cluster expansion of the products of $NN$-correlations, and then
formulate successively HMAMD, TOAMD and TO-HMAMD approaches. 

\subsection{Antisymmetrized Molecular Dynamics (AMD)}
These HMAMD, TOAMD and TO-HMAMD methods are based on the AMD wave function which
is defined as the Slater determinant of $A$-nucleon system,
\begin{equation}\label{eq:amd}
|\Phi_{\rm AMD}\rangle=
\det \{|\phi_{1}(\bm{r}_1) \cdots \phi_{A}(\bm{r}_A)\rangle\}.
\end{equation}
Here, the single-nucleon states $\phi(\bm{r})$ are expressed in the Gaussian wave
packet form with the range parameter $\nu$ and the centroid $\bm{R}$ multiplied by the
spin-isospin component $\chi_{\tau,\sigma}$,
\begin{eqnarray}\label{single-state}
\phi(\bm{r})\propto e^{-\nu(\bm{r}-\bm{R})^2 } \chi_{\tau,\sigma}.
\end{eqnarray}
The Gaussian centroids $\bm{R}$ are usually determined by the cooling process as
discussed in Refs.~\cite{kanada03,kanada12}. For $s$-shell nuclei ${}^{3}$H and
${}^{4}$He, the centroids are optimized to be $\bm{R}=\bm{0}$, which is obtained in the previous studies of TOAMD \cite{myo17a}, and the corresponding
AMD wave functions are reduced to the $s$-wave states. 

\subsection{Description of $NN$ correlations in cluster expansion}
\begin{figure}[t]
    \centering
    \includegraphics[width=\figwidth\textwidth]{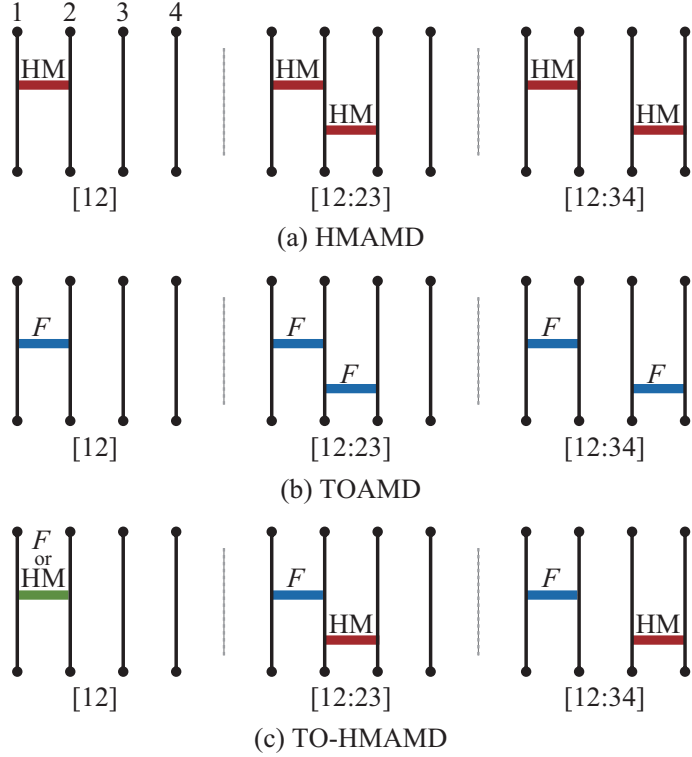}    
    \caption{A part of the correlation diagrams for the $^{4}$He wave functions
        in HMAMD, TOAMD and TO-HMAMD methods. Vertical lines indicate the
        particles, numbering from the left side as 1, 2, 3, 4. The red
        connections entitled ``HM'' denote the $NN$-correlations described by the
        high-momentum pairs of nucleons. The blue connections entitled ``$F$''
        denote the $NN$-correlations described by the TOAMD correlation functions in
        the central or tensor channels. The green connection entitled ``$F$ or
        HM'' stands for either ``$F$" or ``HM''. The label with the square
        bracket below each diagram indicates the configuration of particle
        correlations.}\label{fig:compare}
  \end{figure}

In the AMD wave function $\Phi_{\rm AMD}$, there is no explicit description of
$NN$-correlations, and it is energetically unfavorable in nuclei, when there are
strong tensor force and short-range repulsion in the bare $NN$-interaction.  In
{\it ab initio} calculations, the $NN$-correlations are introduced by
multiplying the correlation functions to the basis state \cite{myo17d}. In the
TOAMD, HMAMD and TO-HMAMD methods, we adopt the AMD wave function as the basis
state and treat the correlations between nucleons in terms of the ``cluster
expansion", where many-body wave function of nuclear system can be expanded into
power series of $NN$-correlations \cite{myo17d,myo18}. In
Fig.~\ref{fig:compare}, we show the first and second orders of correlation
diagrams appearing in the $^{4}$He wave functions corresponding to the HMAMD,
TOAMD and TO-HMAMD methods, respectively. In these diagrams, the
$NN$-correlations are denoted by connections of particle lines and presented in
different colors according to their mathematical descriptions as introduced in
the following paragraphs. The column entitled ``[12]'' contains the two-body
diagrams in which only the nucleons labeled ``1'' and ``2'' are correlated. The
``[12:23]'' denotes the connected three-body diagrams with the double product of
the $NN$-correlations connecting ``[12]" and ``[23]", which belong to the second
order in the cluster expansion. The ``[12:34]'' denotes the disconnected
four-body diagrams with the double product of correlations ``[12]'' and
``[34]''. Some other diagrams, such as the ladder term ``[12:12]'', are not
included in this figure. The cluster expansion in TOAMD approach is explained in
detail in Ref.~\cite{myo17c}. 

To obtain the total wave function of nuclei, we superpose the original AMD wave
function and the correlated bases described by the diagrams in the cluster
expansion, to the first or second order. When only a single $NN$-correlation
such as ``[12]'' is included in HMAMD and TOAMD method, we call the corresponding
methods ``single HMAMD'' and ``single TOAMD'', respectively. When additional
second order diagrams with the double product of $NN$-correlations are also
included, these two methods are named ``double HMAMD'' and ``double TOAMD'',
respectively \cite{myo17c,myo18}. In the TO-HMAMD method, we include at least
first and second order diagrams.

\subsection{High-momentum Antisymmetrized Molecular Dynamics (HMAMD)}
\label{subsec:hmamd}
In the HMAMD method, the $NN$ correlations are described by introducing
high-momentum excitations of nucleon pairs into the AMD wave function, utilizing the
imaginary parts of the Gaussian centroids \cite{myo17e, myo18, itagaki18,
matsuno18}. For $s$-shell nuclei, we denote the excited pairs according to their
spin-isospin combinations:
\begin{equation}\label{eq:pair-kind}
    1: \bm{D}_{p\uparrow,n\uparrow  },\quad
    2: \bm{D}_{p\uparrow,n\downarrow},\quad
    3: \bm{D}_{n\uparrow,n\downarrow  },\quad
    4: \bm{D}_{p\uparrow,p\downarrow  },\quad    
\end{equation}
where the subscripts denote two nucleons in each pair with their spins in the 
$z$-direction. The vector symbol $\bm{D}$ denotes the imaginary shifts of the 
Gaussian centroids for the paired nucleons as
\begin{equation}\label{eq:shift}
    \begin{split}
        \bm{R}_i &\rightarrow \bm{R}_i + i \bm{D},\\
        \bm{R}_j &\rightarrow \bm{R}_j - i \bm{D},
    \end{split}
\end{equation}
where the subscripts $i$ and $j$ denote each nucleon in the $NN$ pair
introduced in Eq.~(\ref{eq:pair-kind}). The vector $\bm{D}$ excites only the
relative motion between two nucleons. Both plus and minus signs are included in
Eq.~(\ref{eq:shift}) to ensure the parity symmetry. The first three cases 1, 2
and 3 in Eq.~(\ref{eq:pair-kind}) should be included in the wave function of the 
${}^{3}$H nucleus, and the ${}^{4}$He nucleus requires additional case 4. The
symmetric spin-isospin states, such as $\bm{D}_{p\uparrow,p\uparrow}$, are
dropped to ensure the total antisymmetrization for the ground states of
$s$-shell nuclei of which the spatial wave functions are symmetric. This
prescription of the correlated pair is extendable to multi-pairs \cite{myo18}.

As discussed in Ref.~\cite{myo17e, myo18}, the imaginary component of the
Gaussian centroid $\textrm{Im}(\bm{Z})$ is proportional to the mean value of
nucleon momentum as $\langle \bm{k} \rangle=2\nu\cdot \textrm{Im}(\bm{Z})$ in
the single-nucleon state described in Eq.~(\ref{single-state}). The finite
$\bm{D}$ in Eq.~(\ref{eq:shift}) corresponds to a high-momentum excitation of
the $NN$ pairs in Eq.~(\ref{eq:pair-kind}), hence they are named
``high-momentum pairs''. In Ref.~\cite{myo17e}, we examined the physical role of
high-momentum pairs in terms of the shell model. It is found that high-momentum
pairs provide the equivalent effect of the full amount of the 2p-2h excitations
described in tensor-optimized shell model frameworks
\cite{myo05,myo07,myo07_11,myo09,myo11}. It is also noted that in
Ref.~\cite{itagaki18}, they consider two-deuteron model of $^{4}$He, where one
nucleon of each deuteron has a high-momentum component.

In single HMAMD calculations, their bases are formulated by introducing a single
high-momentum pair in each AMD basis wave function. The imaginary shift in each
high-momentum pair is described in Eq.~(\ref{eq:shift}) and the vector $\bm{D}$
is selected to be aligned in the spin parallel $z$- and spin perpendicular
$x$-directions \cite{myo17e}. In the double HMAMD calculations, two
high-momentum pairs are additionally introduced in each AMD basis wave function,
as shown in the connected and disconnected cases in Fig.~\ref{fig:compare}. As
an example of two high-momentum pairs for three nucleons, we show the shifted
Gaussian centroids for the connected case \cite{myo17d} as
\begin{equation}\label{eq:shift-double}
    \begin{split}
        \bm{R}_i &\to \bm{R}_i +  i \bm{D}_{1} +  i\bm{D}_{2},\\
        \bm{R}_j &\to \bm{R}_j -  i \bm{D}_{1},\\
        \bm{R}_k &\to \bm{R}_k -  i \bm{D}_{2},
    \end{split}
\end{equation}
where $i$, $j$ and $k$ denote three nucleons with two kinds of the connected
high-momentum pairs $\bm{D}_{1}$ and $\bm{D}_{2}$. For the bases with two
high-momentum pairs, the $z$ and $x$-directions are adopted for the first pair
and $x$, $y$ and $z$-directions are taken for the second pair \cite{myo18}.

Usually, the HMAMD bases are constructed with the good quantum number $K$ for
the $z$-component of total angular momentum. In this case, the rotational
symmetry is restored by the projection of basis $\ket{\Psi_{\rm HMAMD}}$ onto
the eigenstates of the total angular momentum $J$ with the operator
$\hat{P}^J_{MK}$ \cite{schuck80}, as
\begin{equation}\label{eq:ap}
    \begin{split}
    &\left| \Psi^{JM}_{\textrm{HMAMD}, n} \right\rangle\\
    &\quad=\hat{P}_{MK}^{J}\left| \Psi_{\textrm{HMAMD}, n} \right\rangle\\
    &\quad=\frac{2J+1}{8\pi^{2}}\int d \Omega D^{J*}_{MK}(\Omega)\hat R (\Omega)
        \ket{\Psi_{\textrm{HMAMD},n}},
    \end{split}
\end{equation}
where $J$ is the quantum number of the total angular momentum, $K$ is the
magnetic quantum number in the intrinsic state before rotation $\hat R (\Omega)$
of Eular angles $\Omega$, and $M$ is the magnetic quantum number after the
projection. Subscript $n$ denotes all the parameters in the HMAMD basis
including the spin-isospin channels and imaginary shifts $\bm{D}$ for
high-momentum pairs. In numerical calculations, the integration over Euler
angles are performed with the Gauss-Legendre quadrature algorithm. When $K$ is
not a good quantum number for basis $\ket{\Psi_{\textrm{HMAMD},n}}$,
superposition of the basis states with different $K$ could be adopted after the
angular momentum projection, as
\begin{equation}\label{eq:ap2}
\begin{split}
&\left| \Psi^{JM}_{\textrm{HMAMD}, n} \right\rangle\\
&\quad=\sum_{K}c_{K}\hat{P}_{MK}^{J}\left| \Psi_{\textrm{HMAMD}, n} \right\rangle,
\end{split}
\end{equation}
where $c_{K}$ is the coefficient in the superposition, and determined by the
diagonalization of the Hamiltonian matrix with respect to the projected bases
$\hat{P}_{MK}^{J}\left| \Psi_{\textrm{HMAMD}, n} \right\rangle$. 

After the angular momentum projection, HMAMD bases of different spin-isospin
channels and the various imaginary shift vectors $\bm{D}$ are superposed with
the original AMD basis. It is found that the magnitudes of $|\bm{D}|$ ranging
from 1 fm to 12 fm with an interval of 1 fm is sufficient in the superposition
to provide the converging energy in the single HMAMD calculation \cite{myo17e}.
After the superposition, $NN$ correlations in nuclei are described precisely,
including the tensor \cite{myo17e} and short-range correlations \cite{myo18}.

One significant advantage of the HMAMD approach is its simplicity in analytical
derivations of matrix elements. Due to the fact that all HMAMD bases are merely
the Slater determinants of the shifted Gaussians with complex centroids, all the
matrix elements in the HMAMD bases have the same analytical form as those of the
AMD calculation. Hence, only analytical derivations of the AMD matrix elements
are necessary. Meanwhile, the formulation of the HMAMD bases leads to difficulty
in numerical calculation, which is manly caused by the relative angles between
two kinds of vectors for imaginary shifts when two high-momentum pairs are
included simultaneously in the double HMAMD calculation. For this, six bases
corresponding to different combinations of pair directions ($x,z$ for the first
pair and $x,y,z$ for the second pair) should be included in model space for each
combination of magnitudes $|\bm{D}_{1}|$ and $|\bm{D}_{2}|$, which significantly
enlarges the number of bases.

\subsection{Tensor-optimized Antisymmetrized Molecular Dynamics (TOAMD)}
\label{subsec:toamd}
In the TOAMD approach, $NN$ correlation functions are formulated
explicitly in operator forms and then multiplied successively to the AMD basis
\cite{myo15}. In the lowest order, the single TOAMD wave function is written as:
\begin{equation}\label{eq:toamd}    
    (1+F_D+F_S)\times |\Psi_{\rm AMD}\rangle,
\end{equation}
where the operators $F_D$ and $F_S$ correspond respectively to the tensor and
short-range correlations. In practical calculations, they are formulated in the
Gaussian expansion form,
\begin{align}
&F_{D}= \sum_m^{n_{\textrm{G}}} 
    \sum_{t} C_{D,m}^t f_{D,m}^t,\label{eq:f-toamd}\\
&F_{S}= \sum_m^{n_{\textrm{G}}} 
    \sum_{t,s} C_{S,m}^{t,s} f_{S,m}^{t,s},\label{eq:f-toamd-2}
\end{align}
where
\begin{align}
&f_{D,m}^t= \sum_{i<j}^A \exp(-a^t_{D,m} r_{ij}^2) 
    O_{ij}^t  r_{ij}^{2} S_{12}(\hat{r}_{ij}),\label{eq:fd}\\
&f_{S,m}^{t,s}= \sum_{i<j}^A \exp(-a^{t,s}_{S,m} r_{ij}^2)
    O_{ij}^t O_{ij}^s. \label{eq:fs}
\end{align}
Here, $s$ and $t$ are used to represent the spin-isospin channels of the
correlated two nucleons and $O_{ij}^t=(\bm \tau_i \cdot \bm \tau_j)^t$,
$O_{ij}^s=(\bm \sigma_i \cdot \bm \sigma_j)^s$. The vector
$\bm{r}_{ij}=\bm{r}_i-\bm{r}_j$ is the relative coordinate between the correlated
two nucleons. The subscript $m$ denotes the different range parameters
$a^t_{D,m}$ and $a^{t,s}_{S,m}$ in the Gaussian expansion of each channel with
the number of $n_G$. The properties of these correlation functions are discussed
in Refs.~\cite{myo17b,myo17c}. 

In double TOAMD calculations, up to the second order of diagrams are included by
using double product of the correlation functions $F_{S}$ and $F_{D}$, as
\begin{equation}\label{eq:toamd2}
    \begin{aligned}
    (1+F_D+F_S +F_S F_S +F_S F_D
       +F_D F_S +F_D F_D)\times |\Psi_{\rm AMD}\rangle.
    \end{aligned}    
    \end{equation}
Parameters in each correlation function $F$ are determined independently and
have different coefficients. Comparing with the double HMAMD method, the double
TOAMD approach converges faster to the exact solution with the small number of
bases. However, this numerical efficiency relies on the enormous efforts to
obtain the analytical derivations of matrix elements coming from the double
product of $FF$ in Eq.~(\ref{eq:toamd2}). For instance, when calculating the
two-body matrix element $\braket{\Phi_{\rm AMD} |F^\dagger F^\dagger
VFF|\Phi_{\rm AMD}}$ for $^{4}$He, there are 336 diagrams of up to the four-body
terms in the cluster expansion of many-body operator $F^\dagger F^\dagger VFF$
\cite{myo17c}. For each of these diagrams, the analytical formulation of matrix
elements needs to be derived. In addition, it takes efforts for the code
development.

In Ref.~\cite{myo18}, the equivalent results between the correlation functions
$F$ and high-momentum pairs are discovered. We compared numerical results from
double HMAMD and double TOAMD calculations using the AV4$^\prime$ central
interaction with short-range repulsions. In Appendix, we show explicitly the
analytical proof for the equivalence between high-momentum pairs and correlation
functions $F$ for the central type of $NN$-correlation under some specific
conditions.
  
\subsection{Tensor-optimized High-momentum Antisymmetrized Molecular Dynamics
(TO-HMAMD)} 

\begin{figure}[t]
    \centering
    \includegraphics[width=\figwidth\textwidth]{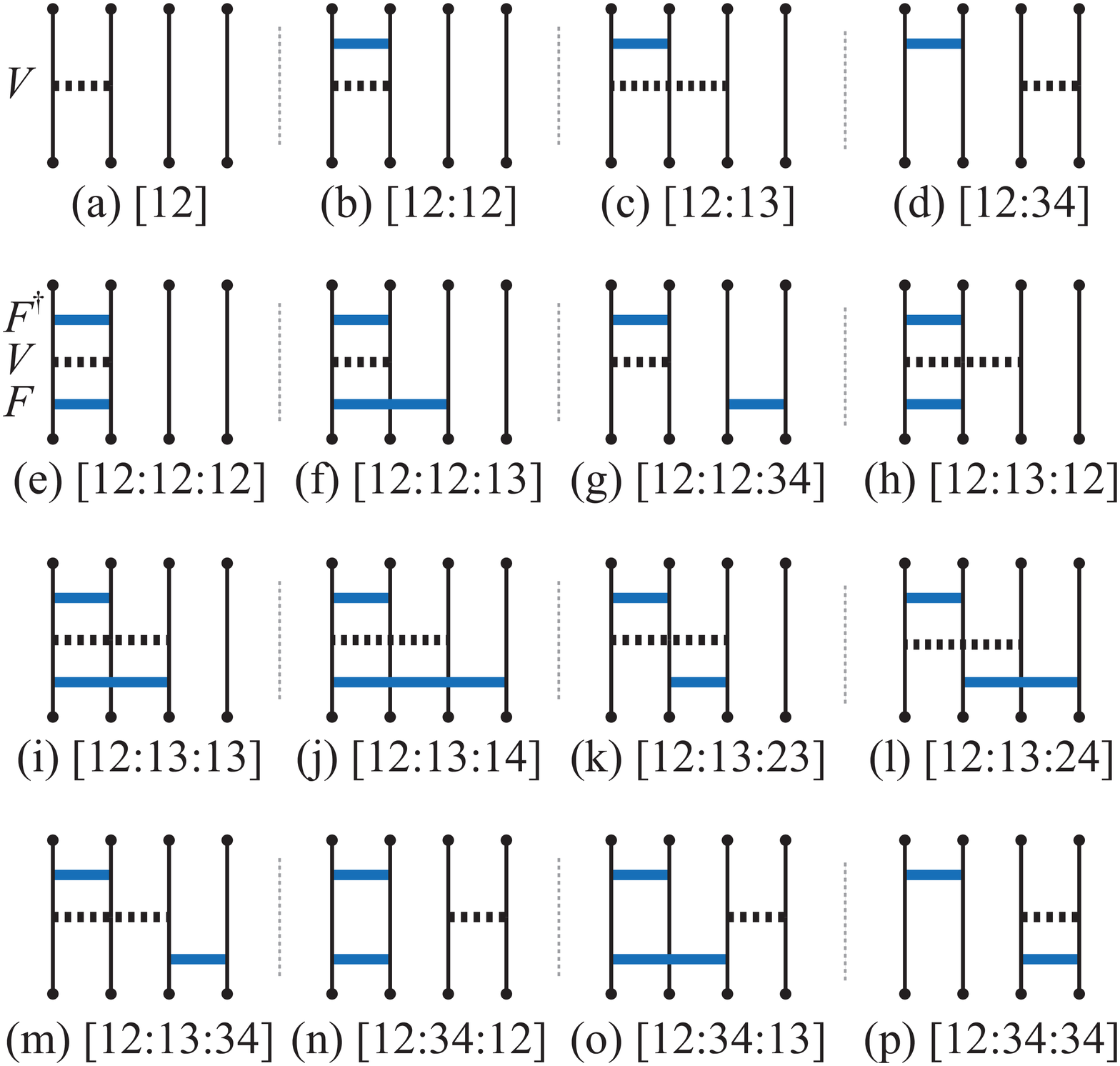}
    \caption{Diagrams of cluster expansions in calculating matrix elements for the 
    two-body operator $V$. The diagram (a) is for the operator $V$. The diagrams (b-d)
    correspond to the cluster expansion of the operator $F^{\dagger}V$. The diagrams (e-p)
    correspond to the cluster expansion of the operator $F^{\dagger}VF$.  The blue
    connections indicate the TOAMD correlation functions $F_D$ or $F_S$. The dotted
    connections indicate the two-body operator $V$. For each diagram, the
    corresponding integral kernels should be derived analytically. }
    \label{fig:diagram}
\end{figure}

In order to integrate advantages from both HMAMD and TOAMD methods, and balance
between the analytical simplicity and numerical efficiency in {\it ab initio}
calculations, we propose the TO-HMAMD method by hybridizing the HMAMD and TOAMD
descriptions of $NN$-correlations. The correlation diagrams for the TO-HMAMD
approach are shown in the bottom panel of Fig.~\ref{fig:compare}. For the first
order of diagrams, such as ``[12]'', the $NN$-correlations are described by
using either the high-momentum pair with the imaginary shift $\bm{D}$ or the
TOAMD correlation function $F_{S}$ or $F_D$. For the second order diagrams, the
$NN$ correlations are described in combination of both the high-momentum nucleon
pair and the TOAMD correlation function $F$. The TO-HMAMD wave function can be
expressed in the following form as
\begin{align}
|\Psi_{\rm TO-HMAMD}^{JM} \rangle 
&=  \sum_{n} C_{n}
    (1+F_{D,n}+F_{S,n})\times \sum_{K}c_{K} \hat{P}^J_{MK}|\Psi_{{\rm HMAMD},n}\rangle, \nonumber \\
&=      \sum_{\mu}  {\tilde C}_{\mu}\:
    \hat{P}^J_{MK_{\mu}} {\tilde F}_{\mu}  |\Psi_{{\rm HMAMD},\mu}\rangle \nonumber\\ 
&=  \sum_{\mu}  {\tilde C}_{\mu}\: 
    \hat{P}^J_{MK_{\mu}} \ket{\Psi_{{\rm TO-HMAMD},\mu}},    \label{eq:to-hmamd}
\end{align}
where $\ket{\Psi_{{\rm HMAMD},n}}$ are HMAMD bases with a single high-momentum
pair and $F_{D}$ and $F_{S}$ are the correlation functions in
Eq.~(\ref{eq:f-toamd}). The operator $\hat{P}^{J}_{MK}$ represents the angular
momentum projection and $c_{K}$ is the corresponding superposition coefficients
for the quantum number $K$, as in Eq.~(\ref{eq:ap2}). For the study of the
$^{4}$He nucleus in the present work, the HMAMD bases $\ket{\Psi_{{\rm
HMAMD},n}}$ are constructed with the good quantum number $K$ in the intrinsic
frame, and the summation over $K$ reduces to a single term with $K=0$.
Subscripts $n$ for the correlation functions $F_D$ and $F_S$ indicate that
parameters in $F_{D,n}$ and $F_{S,n}$ are determined independently for each
HMAMD basis $\ket{\Psi_{{\rm HMAMD},n}}$. The operator ${\tilde F}_{\mu}$ in the
second line is selected from the operators $\{1, f_{D,m}^{t}, f_{S,m}^{t,s}\}$
that correspond to three different channels where $f_{D,m}^{t}$ and
$f_{S,m}^{t,s}$ are defined in Eqs.~(\ref{eq:fd}) and (\ref{eq:fs}). The
subscripts $\mu$ denote all the adjustable parameters, including the quantum
number $K_{\mu}$, the spin-isospin combination and the imaginary shift $\bm{D}$
(including $\bm{0}$) of high-momentum pairs in $\ket{\Psi_{{\rm HMAMD},n}}$, and
the quantum numbers for spin $s$ and isospin $t$, and the Gaussian range
parameter $m$ of the operator ${\tilde F}$. $\ket{\Psi_{{\rm TO-HMAMD}, \mu}}=
{\tilde F}_{\mu} |\Psi_{{\rm HMAMD}, \mu}\rangle$ are defined as TO-HMAMD bases
and ${\tilde C}_\mu$ are the corresponding expansion coefficients. Then, the
energy and wave function of a nucleus could be obtained by solving the
Hill-Wheeler equation, as
\begin{equation}\label{eq:hill-wheeler}
    \sum_{\mu,\mu'}\left( H_{\mu \mu'}-EN_{\mu
        \mu'}\right) \tilde C_{\mu'} =0, 
\end{equation}
where $H_{\mu\mu'}$ and $N_{\mu\mu'}$ are the Hamiltonian and norm matrix
elements, respectively, as
\begin{align}
H_{\mu \mu'}&=\langle \Psi_{\rm TO-HMAMD,\mu} 
    |\hat{P}^{J\dagger}_{MK_{\mu}}H \hat{P}^J_{MK_{\mu'}}| \Psi_{\rm TO-HMAMD,\mu'}\rangle, \\
N_{\mu \mu'}&=\langle \Psi_{\rm TO-HMAMD,\mu} 
    |\hat{P}^{J\dagger}_{MK_{\mu}}\hat{P}^J_{MK_{\mu'}}|   \Psi_{\rm TO-HMAMD,\mu'}\rangle.
\end{align}
In the above matrix elements, the integrations over the Euler angles $\Omega$ in
$\hat{P}^J_{MK_{\mu}}$ are performed numerically but the integral kernels such as
$\braket{\Psi_{\rm TO-HMAMD,\mu}| H |\Psi_{\rm TO-HMAMD,\mu'}}$ are obtained
analytically. As discussed in Subsection~\ref{subsec:hmamd}, the introduction of
high-momentum pairs does not require the additional components of the analytical
forms of matrix elements. Hence, in analytical derivations of the matrix
elements, $\braket{\Psi_{\rm TO-HMAMD,\mu}|H|\Psi_{\rm TO-HMAMD,\mu'}}$ is
reduced to the matrix element $\braket{\Psi_{\rm TOAMD,\mu}|H|\Psi_{\rm
TOAMD,\mu'}}$ for the single TOAMD wave function, which is to be obtained in term of
the cluster expansion \cite{myo17c}.  As an example, in Fig.~\ref{fig:diagram} we
show 16 diagrams in the cluster expansion associated to the two-body operator $V$, in
the order of $V$ (a), $F^{\dagger}V$ (b-d) and $F^{\dagger}VF$ (e-p) cases.
The integral kernels corresponding to each diagram should be derived analytically.
    
We get now three methods for the treatment of the correlations among nucleons.
In Table \ref{tab:compare}, we compare the analytical and numerical efforts in
detail for the new TO-HMAMD method as compared with the double HMAMD and double
TOAMD approaches. The second row lists the numbers of diagrams in the cluster
expansions that need to be derived analytically in the calculation of matrix
elements for the two-body operator $V$. In the double HMAMD approach,
high-momentum pairs do not change the analytical form of matrix element.
Therefore, all the diagrams are reduced to the most simple type as shown in
Fig.~\ref{fig:diagram}~(a), and only one integral kernel for operator $V$ is to
be derived analytically. In the double TOAMD method, the number of diagrams is
412, including 16 diagrams shown in Fig.~\ref{fig:diagram}, 60 diagrams for
$F^{\dagger}VFF$ and 336 diagrams for $F^{\dagger}F^{\dagger}VFF$, as discussed
in Ref.~\cite{myo17c}. The integral kernel corresponding to each of these
diagrams needs to be derived independently. In the new hybridized TO-HMAMD
method, the diagrams are reduced to 16 different single TOAMD diagrams as shown
in Fig.~\ref{fig:diagram}, which significantly simplifies the analytical
derivation, comparing with the double TOAMD case. It is therefore clearly shown
that in the calculation of the two-body operator $V$, TO-HMAMD requires more
analytical efforts comparing to double HMAMD, but provides significantly simpler
mathematical framework than that of double TOAMD.

\begin{table}[t]
    \begin{center}
    \caption{The comparison of analytical and numerical settings between the double
        HMAMD, double TOAMD and TO-HMAMD methods in the calculation of $^4$He
        ($0^+$). In the first row, we compare the number of diagrams in
        calculating matrix elements for the two-body operator $V$. For each diagram,
        the corresponding integration kernels should be derived analytically. In
        the second and third rows, we compare the number of bases in the
        superposition and the number of mesh points in numerical integrations for the 
        angular momentum projection (AP). The numbers of bases for double HMAMD and
        double TOAMD are the estimated ones. The definitions of the diagrams and bases
        are explained in the text and in Fig.~\ref{fig:diagram}.}
    \label{tab:compare}
    \vspace{2mm}
    \begin{tabular}{rlll}
\noalign{\hrule height 0.5pt}
            &Double~~   		&Double~~     &TO-\\
            &HMAMD~~  		&TOAMD~~      &HMAMD\\
\noalign{\hrule height 0.5pt}
Diagrams Number~~ & $1$~~ 	        &$412$ \cite{myo17c}     & $16$\\
Bases Number~~	    & $13824$~~   &$1369$~~	& $1813$\\
AP mesh points Number~~ &$8000$ \cite{myo18}    &$-$~~         & $14$ \\
\noalign{\hrule height 0.5pt}
    \end{tabular}
    \end{center}
\end{table}

Another advantage in the TO-HMAMD approach is that only the $z$-direction of the
imaginary shift $\bm{D}$ is necessary in the HMAMD part \cite{lyu18}. This is
because of the fact that the relative coordinates $\bm{r}$ between the
correlated two nucleons is integrated over the entire space by using the
correlation functions of $F_D$ and $F_S$ in Eqs.~(\ref{eq:f-toamd}) and
(\ref{eq:f-toamd-2}), which contribute to the high-momentum excitations for the
$NN$-pairs in any direction. Hence, all the possible relative angles between the
two kinds of the correlated $NN$-pairs, described respectively by the
high-momentum pairs and $F_{D,S}$, are naturally taken into account.
Consequently, the number of bases are significantly smaller in the TO-HMAMD
method comparing to the double HMAMD approach, where different directions of
momenta in the first and second high-momentum pairs should be taken into
account. 

In the third row of Table \ref{tab:compare}, we compare the number of bases that
is required in the numerical calculations among the TO-HMAMD, double HMAMD and
double TOAMD methods. The basis number for the double HMAMD calculation is
estimated as follows. We set $n_\textrm{D}$=12 with different magnitudes for the
imaginary shifts $\bm{D}$ in two directions ($x$ and $z$) for both the first and
second pairs and select two sets among four kinds of spin-isospin combinations
in Eq.~(\ref{eq:pair-kind}). The parity doublets for these pairs are also
included. The total base number is then estimated to be $(n_\textrm{D}\times
2\times 2)^2 \times C(4,2)=13824$. For the double TOAMD method, we typically
include $n_\textrm{G}$=6 of different ranges in Gaussian expansion  for each of
the 6 channels of the correlation function $F$ in Eqs.~(\ref{eq:f-toamd}) and
(\ref{eq:f-toamd-2}), and then the total number of bases could be approximated
by $(1+6\times n_\textrm{G})^2=1369$. In TO-HMAMD, we adopt all the $4$ possible
spin-isospin combinations in Eq.~(\ref{eq:pair-kind}) for each high-momentum
pair in the HMAMD bases $\ket{\Psi_\textrm{HMAMD}}$ in Eq.~(\ref{eq:to-hmamd}).
Each pair is fixed in the $z$ direction with $n_\textrm{D}=6$ different
magnitudes of the imaginary shift and their parity doublets are also included.
In the correlation functions $F$ in Eq.~(\ref{eq:to-hmamd}), we include for each
of the 6 channels $n_\textrm{G}=6$ proper ranges for the Gaussian expansion.
Hence the total number of TO-HMAMD bases is $(1+4\times 2\times
n_\textrm{D})\times(1+6\times n_\textrm{G})$=1813. It is observed that the
number of bases in TO-HMAMD is similar to that in double TOAMD but much smaller
than that in double HMAMD method. From this fact, we conclude that the numerical
efficiencies of TO-HMAMD and double TOAMD in describing the exact wave function
are better as compared to the double HMAMD method.

In the application to the $0^+$ ground state of ${}^{4}$He, there is one
additional advantage in the TO-HMAMD approach. In this state, the TO-HMAMD wave
function has the rotational symmetry around the $z$-axis and hence the angular
momentum projection operator $\hat{P}^J_{MK}$ in Eq.~(\ref{eq:to-hmamd}) reduces
to
\begin{equation}\label{eq:ap-beta}
    \hat{P}_{00}^{0}
        =\frac{1}{8\pi^{2}}\int d \beta \sin\beta \hat R (\beta),
\end{equation}
where the rotation over the Eular angles $\Omega=\{\alpha,\beta,\gamma\}$ is reduced
to a single rotation over the polar angle $\beta$. This reduces significantly the
numerical efforts because the integration over the angle $\alpha$ and $\gamma$ can be
skipped. In the TO-HMAMD calculation of $^{4}$He($0^+$), we adopt 14 mesh points in
the numerical integration over the angle $\beta$. This reduction of the effort for the
angular momentum projection is only valid for the calculation of the $0^+$
states and it is not available for the calculation of $^{3}$H nucleus, the ground
state of which has the spin-parity $1/2^+$. In addition, this reduction relies on
the rotational symmetry around $z$-axis, which is broken in the double HMAMD
wave function by the high-momentum pairs in the $x$-direction. Hence, in this case the 
numerical integration has to be carried out for all of the three Euler angles. In
the previous double HMAMD calculation, the total number of mesh points for the 
projection is $20\times20\times20=8000$ \cite{myo18}. In the fourth row of Table
\ref{tab:compare}, we compare the total number of mesh points for the angular
momentum projection between the double HMAMD and TO-HMAMD calculations, which
further illustrate the numerical efficiency in the TO-HMAMD approach. In addition,
we note that the double TOAMD wave function has the intrinsic spin-parity $0^+$ for
the ground state of $^{4}$He, because of the scalar nature of the correlation
functions $F$ in Eqs.~(\ref{eq:f-toamd}) and (\ref{eq:f-toamd-2}). Therefore, the 
angular momentum projection is not necessary for the double TOAMD method. 

As a balance between the analytical and numerical simplicity, the TO-HMAMD
approach can be extended to the $p$-shell nuclei with minimal efforts, which
makes it a promising {\it ab initio} method in the $p$-shell nuclear system.

\section{Results for the ${}^{4}$H\lowercase{e} nucleus}
\label{sec:s-shell}
\begin{figure}[b]
    \centering 
    \includegraphics[width=\figwidth\textwidth]{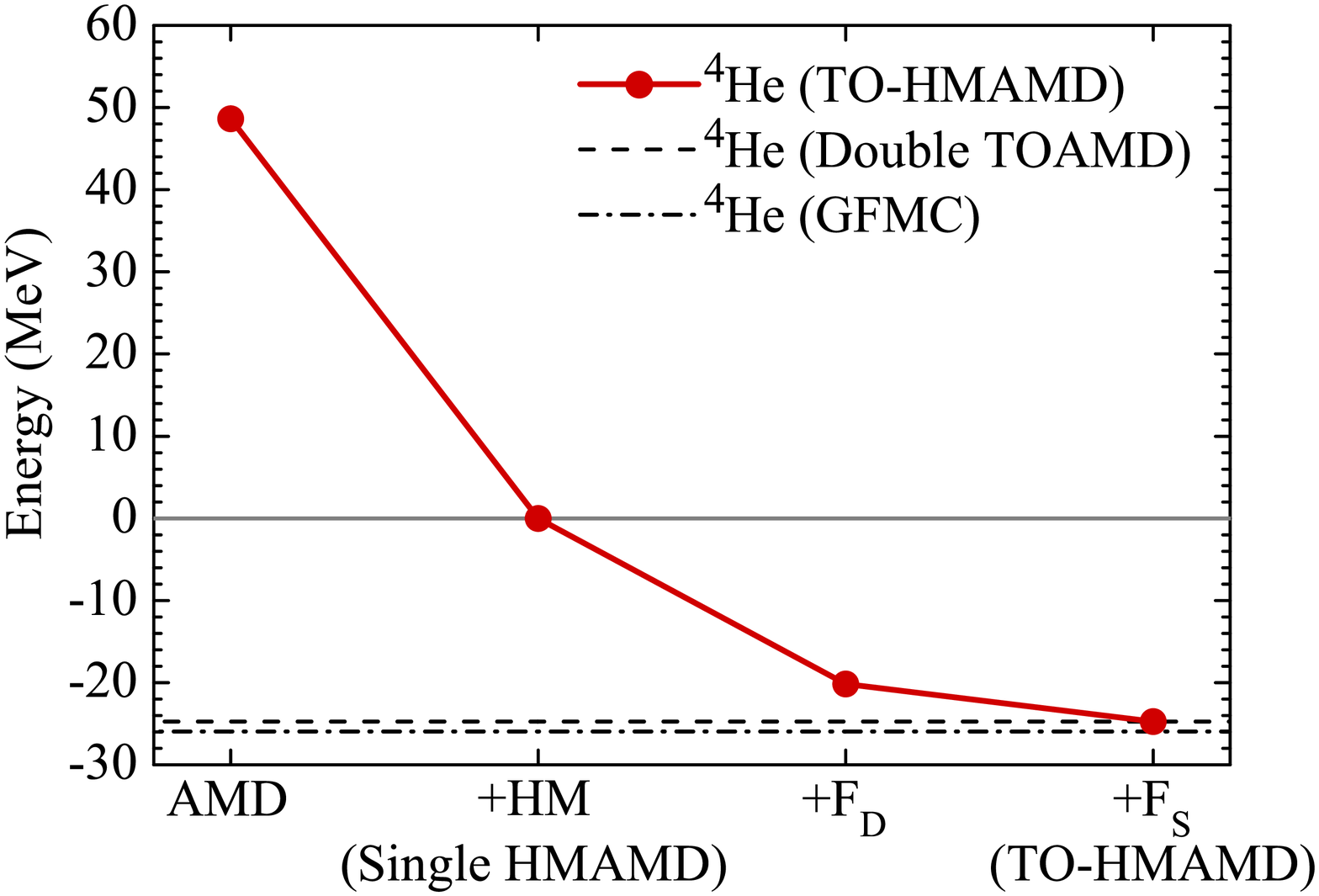}
    \caption{\label{fig:energy-4He} The energy of the ${^4}$He nucleus calculated
    with TO-HMAMD using the bare interaction AV8$^\prime$ by successively adding
    the high-momentum pairs (+HM), and correlation functions $F_D$ and $F_S$.} 
\end{figure}

We perform the {\it ab initio} calculation of the ${}^{4}$He nucleus with the
TO-HMAMD method using the AV8$^\prime$ bare interaction. The Gaussian range
parameter $\nu$ in Eq.~(\ref{eq:amd}) is variationally optimized as $\nu$=0.20
fm$^{-2}$. The corresponding energy curve of ${}^{4}$He by successively adding
high-momentum pairs, and correlation functions $F_D$ and $F_S$ are presented in
Fig.~\ref{fig:energy-4He}. It is observed that both the additions of
high-momentum pairs and correlation functions $F_{S,D}$ significantly improve
the total energy of $^{4}$He, and the final TO-HMAMD result converges with the
double TOAMD calculation. 

\begin{figure}[t]
    \centering 
    \includegraphics[width=\figwidth\textwidth]{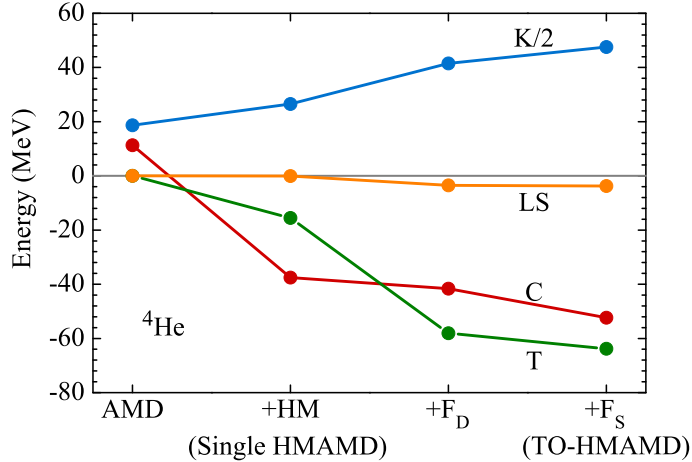}
    \caption{\label{fig:components-4He} Hamiltonian components of the ${^4}$He
    nucleus calculated with TO-HMAMD using the bare interaction AV8$^\prime$ by
    successively adding the high-momentum pairs (+HM), and correlation functions
    $F_D$ and $F_S$. ``K/2'' denotes a half of kinetic component. ``C'', ``T''
    and ``LS'' denote central, tensor and spin-orbit components, respectively.} 
\end{figure}

\begin{table}[t]
    \begin{center}
    \caption{Total energies, Hamiltonian components and root-mean-square radii
        of $^4$He ($0^+$) calculated with TO-HMAMD using the bare interaction
        AV8$^\prime$ by successively adding the high-momentum pairs (+HM), and
        correlation functions $F_D$ and $F_S$. The units of energies and radii
        are MeV and fm, respectively.}
    \label{tab:evolution}
\begin{tabular}{crrrr}
\noalign{\hrule height 0.5pt}
~~~&AMD~~&$+\textrm{HM}~~~$&~$+F_\textrm{D}$~~~  &~~$+F_\textrm{S}$~~~\\
\noalign{\hrule height 0.5pt}
~~~Energy 	& ~$48.64$~~~     & $-0.01$~~~~	& $-20.14$~~	&$-24.74$\\
~~~Kinetic	& ~$37.32$~~~	    & $ 53.11$~~~~	& $82.99$~~	    &$ 95.17$\\
~~~Central	& ~$11.31$~~~	    & $-37.50$~~~~	& $-41.61$~~	&$-52.33$\\
~~~Tensor 	& ~$0.00$~~~	    & $-15.55$~~~~	& $-58.05$~~	&$-63.80$\\ 
~~~LS   	& ~$0.00$~~~ 	    & $-0.07$~~~~ 	& $-3.47$~~	    &$-3.77$\\
\noalign{\hrule height 0.5pt} 
~~~Radius	&  1.68~~   	&~1.71~~~~ 	& 1.58~~	&1.51\\ 
\noalign{\hrule height 0.5pt}
    \end{tabular}
    \end{center}
\end{table}

In Fig.~\ref{fig:components-4He} and Table \ref{tab:evolution}, we show the
evolution of the Hamiltonian components and the radius of $^{4}$He with the successive
addition of high-momentum pairs and correlation functions. We found that both
the central (red curve) and tensor (green curve) terms are improved by the first
inclusion of high-momentum pairs, especially for the central term where the
improvement is as large as 48.8 MeV. This means that both the tensor and
short-range correlations are treated by using high-momentum pairs. It is
interesting that the spin-orbit term remains almost 0 MeV for the AMD case,
showing that the spin-orbit correlation can only be well described by the second
order of correlation diagrams. In the next introduction of tensor correlation
function $F_D$, the tensor term is more significantly improved by 42.5 MeV
comparing to the central term. In addition, contribution of the spin-orbit
term arises with a finite value $-$3.46 MeV as expected, because of the coupling
between the high-momentum pair and tensor correlation function $F_D$. The addition
of the last correlation function $F_S$ contributes to all the Hamiltonian
components. For each addition of high-momentum pairs and correlation functions,
we observe significant increases of the kinetic energy, which correspond to the
high-momentum excitations induced by the short-range repulsion and tensor
attraction in the AV8$^\prime$ bare interaction.

\begin{table}[h]
    \begin{center}
    \caption{ Total energies, Hamiltonian components and root-mean-square radius
    of $^4$He ($0^+$) calculated with TO-HMAMD using AV8$^\prime$ potential in
    comparison with the TOAMD and GFMC methods. ``$F^2$-TOAMD'' denotes the double TOAMD
    method. The units of energies and radii are MeV and fm, respectively.}
    \label{tab:4He}
    \vspace{4mm}
\begin{tabular}{crrr}
\noalign{\hrule height 0.5pt}
~~~&TO-HMAMD  		&~$F^2$-TOAMD\cite{myo17c}   &~GFMC\cite{kamada01}~~~~~~\\
\noalign{\hrule height 0.5pt}
~~~Energy 	& $-24.74$~~~~~ & $-24.74$~~~~~	& $-25.93$~~~~~	\\
~~~Kinetic	& $ 95.17$~~~~~	& $ 97.06$~~~~~	& $102.3$~~~~~	\\
~~~Central	& $-52.33$~~~~~	& $-53.12$~~~~~	& $-55.05$~~~~~	\\
~~~Tensor 	& $-63.80$~~~~~	& $-64.84$~~~~~	& $-68.05$~~~~~	\\ 
~~~LS   	& $-3.77$~~~~~ 	& $-3.83$~~~~~ 	& $-4.75$~~~~~	\\
\noalign{\hrule height 0.5pt} 
~~~Radius	&  1.51~~~~~   	&~1.50~~~~~ 	& 1.49~~~~~	\\ 
\noalign{\hrule height 0.5pt}
    \end{tabular}
    \end{center}
\end{table}

We compare the TO-HMAMD results of the $^{4}$He nucleus with the double TOAMD and
GFMC methods by listing the total energy, Hamiltonian components and the
root-mean-square radius obtained from each method in Table \ref{tab:4He}. Nice
agreements are found for each component between the TO-HMAMD and other two
methods. The TO-HMAMD method is found to reproduce exactly the same total energy
of $^{4}$He nucleus as the double TOAMD calculation, showing that these two
methods describe almost the same wave function. Comparing with the double TOAMD
results, the kinetic energy in TO-HMAMD results is slightly smaller. This
indicates that the high-momentum components are slightly underestimated in the
current calculation. When the TO-HMAMD calculation is performed with larger
model space, it is expected that small difference between current result and
precise solution can be further reduced.

\begin{figure}[b]
    \centering
    \includegraphics[width=\figwidth\textwidth]{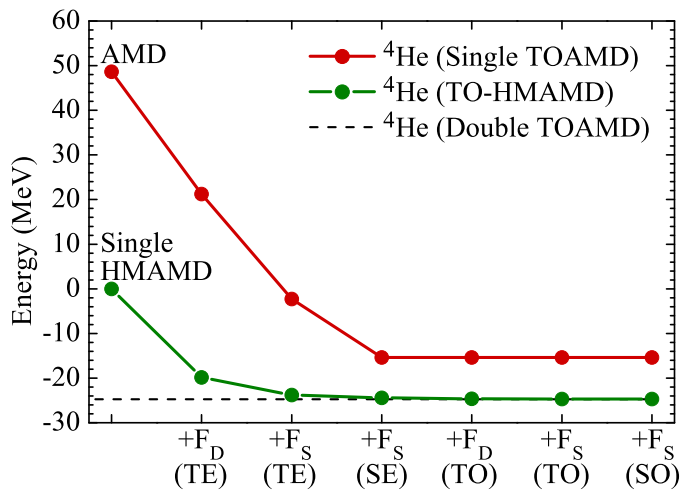}
    \caption{The energy convergence of $^{4}$He with respect to the spin-isospin
    channels of the correlation functions $F_{D}$ and $F_S$ in TO-HMAMD and
    single TOAMD calculations.}\label{fig:helium-channel}
\end{figure}

\section{Spin-isospin channel dependence of the $NN$-correlation}
\label{sec:corr}
\begin{figure}[t]
    \centering
    \includegraphics[width=\figwidth\textwidth]{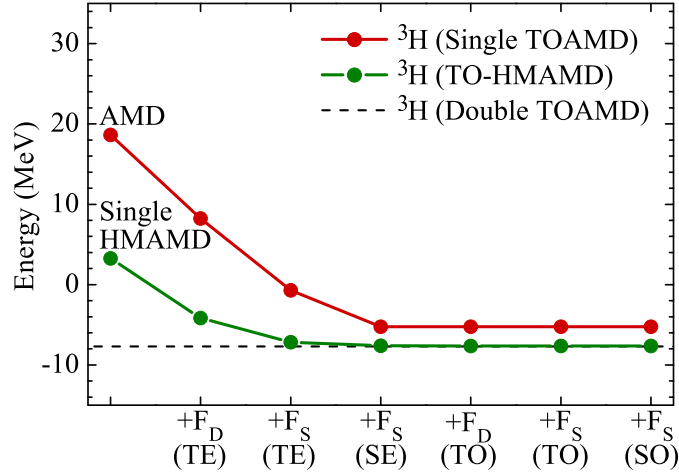}
    \caption{The energy convergence of $^{3}$H with respect to the spin-isospin
    channels of the correlation functions $F_{D}$ and $F_S$ in TO-HMAMD and
    single TOAMD calculations.}\label{fig:triton-channel}
\end{figure}

We further discuss the spin-isospin channel dependence of $NN$-correlations in
both descriptions using high-momentum pairs or correlation functions $F$. In
Fig.~\ref{fig:helium-channel}, we show the spin-isospin channel dependence for
the $NN$-correlations described by the correlation function $F$ in the single
TOAMD and TO-HMAMD calculations of $^{4}$He.  In this figure, we project the
correlation functions $F_{S}$ and $F_{D}$ in Eqs.~(\ref{eq:f-toamd}) and
(\ref{eq:f-toamd-2}) into spin-isospin eigenstates of the correlated two
nucleons, and add these correlation functions successively in the order of
triplet-even (TE), singlet-even (SE), triplet-odd (TO) and singlet-odd (SO)
channels. As shown by the red curve in Fig.~\ref{fig:helium-channel}, which
corresponds to the single TOAMD calculation, the even channels contribute to the
entire energy improvements from the AMD basis, while the odd channels have
exactly no effect. This originates from the total even parity of the $s$-wave
AMD basis state of ${}^{4}$He, and single correlation function $F$ should be in
the even channel to preserve the parity of the AMD basis state. In the case of
the TO-HMAMD calculation (green curve), we observe a similar dominance of the
even channels. However, in this case, contributions from the odd channels are
very small but finite, because of the coupling terms between odd TOAMD
correlation functions and odd channels of high-momentum pairs. Furthermore, the
triplet-even channel contributes to about 95\% of the total energy improvement,
which agrees with the prediction from the one-pion-exchange process
\cite{carlson98}. In Fig.~\ref{fig:triton-channel}, we show similar channel
dependence of the correlation functions $F$ in the ${}^{3}$H nucleus where total
energy is obtained as $-$7.64 MeV with the AV8$^\prime$ interaction.

\begin{figure}[t]
    \centering
    \includegraphics[width=\figwidth\textwidth]{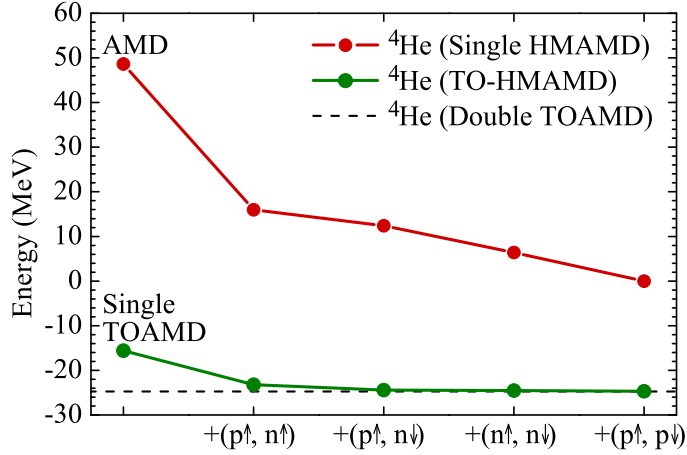}
    \caption{Energy convergence of $^{4}$He with respect to the spin-isospin
        combinations of the high-momentum pairs in TO-HMAMD and single HMAMD
        calculations.}\label{fig:helium-pair}
\end{figure}

\begin{figure}[t]
    \centering
    \includegraphics[width=\figwidth\textwidth]{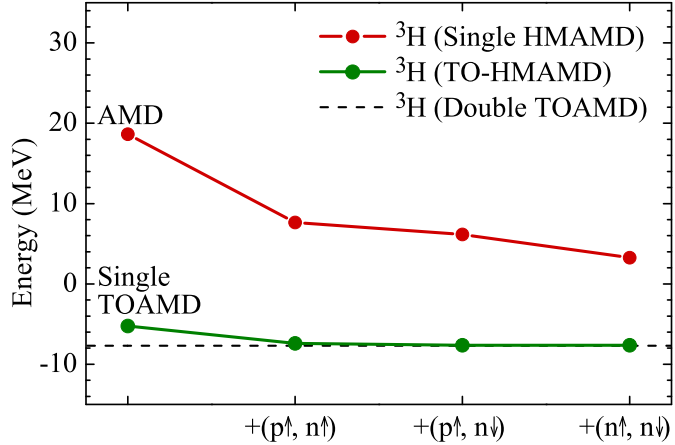}
    \caption{Energy convergence of $^{3}$H with respect to the spin-isospin
        combinations of the high-momentum pairs in TO-HMAMD and single HMAMD
        calculations.}\label{fig:triton-pair}
\end{figure}

The $NN$-correlation is also described by the high-momentum pairs of nucleons in the 
TO-HMAMD calculation. In Fig.~\ref{fig:helium-pair}, we show the energy
contribution from each successive addition of various spin-isospin combinations
for the high-momentum pairs in Eq.~(\ref{eq:pair-kind}). It is found that in
both single HMAMD and TO-HMAMD calculations, the $(p\uparrow,n\uparrow)$
high-momentum pairs contribute to the most of the energy improvements comparing with
the original AMD or the single TOAMD basis. On the other hand,
$(n\uparrow,n\downarrow)$ and $(p\uparrow,p\downarrow)$ pairs have a smaller
contribution to the total energy, especially in the TO-HMAMD calculations.
Considering that the $(p\uparrow,n\uparrow)$ pair contains both the triplet-even
and triplet-odd channels, while $(n\uparrow,n\downarrow)$ and
$(p\uparrow,p\downarrow)$ pairs contain only triplet-odd but no triplet-even
channels, we may conclude that the triplet-even channel also dominates the
$NN$-correlations in the description using the high-momentum pairs, as the
previous discussions for the correlation functions $F$. The similar spin-isospin
dependence is also shown for the $^{3}$H nucleus in Fig.~\ref{fig:triton-pair}.
We also notice that the contributions from the $(n\uparrow,n\downarrow)$ and
$(p\uparrow,p\downarrow)$ pairs are finite in Fig.~\ref{fig:triton-pair} instead
of the negligible results reported in Ref.~\cite{myo17e}. This is because of the
fact that the $(n\uparrow,n\downarrow)$ and $(p\uparrow,p\downarrow)$ pairs
improve descriptions of short-range correlations induced by the AV8$^\prime$
interaction in this calculation, while for the calculation in
Ref.~\cite{myo17e}, no short-range repulsion is included in the central
interaction.


In future {\it ab initio} calculations with the TO-HMAMD method, the information
of the dependence on the spin-isospin channel can be utilized for more effective
description of $NN$-correlations in nuclei.

\section{Conclusion}
\label{sec:conclusion} 
In conclusion, we propose the variational ``TO-HMAMD'' framework for {\it ab
initio} calculation of nuclei by hybridizing the tensor-optimized and
high-momentum AMD approaches (TOAMD and HMAMD). The wave function of nuclei is
formulated by using the AMD reference state and the additional $NN$-correlations
expressed in the form of cluster expansion.  The correlation diagrams are
included up to the second order and described by the product of high-momentum
pairs and spatial correlation functions. Comparing with other two methods based
on AMD (double HMAMD and double TOAMD), it is found that the newly proposed
TO-HMAMD approach has an advantage in balancing analytical simplicity and
numerical efficiency, which is ideal for the future extensions to $p$-shell
nuclei. Through the TO-HMAMD approach, the ${}^{4}$He nucleus is calculated
using the AV8$^\prime$ bare interaction. It is found that the total energy,
Hamiltonian components and root-mean-square radius of the ${}^{4}$He nucleus are
well reproduced comparing to the double TOAMD and GFMC results. We also discuss
the spin-isospin channel dependence of $NN$-correlations in TO-HMAMD, and
observe the dominance of even channels, especially the triplet-even channel, in
the description of $NN$ correlations within the ground state of ${}^{4}$He. In
future studies, the application of the TO-HMAMD approach will be extended to
$p$-shell and heavier nuclear systems. Because of its flexibility in describing
the $NN$-correlations and its advantages in analytical and numerical efficiency,
it is expected that the TO-HMAMD can be used as a general {\it ab initio}
framework for the $p$-shell nuclear systems.

\section*{Acknowledgments}
M.L. acknowledges the support from the RCNP theoretical group for his stay in
RCNP and the fruitful discussions with the members, and the support from the
Yozo Nogami Research Encouragement Funding. This work was supported by the JSPS
KAKENHI Grants No. JP18K03660, No. JP15K05091, No. JP15K17662, and No.
JP16K05351. One of the authors (M.I.) is supported by the Grants-in-Aid for
Young Scientists (B) (No. 15K17671) and Grant-in-Aid for JSPS Research Fellow
(No. 16J05297). The numerical calculations were performed on the high
performance computing server at RCNP, Osaka University.

\appendix*
\section{Comparison between HMAMD and TOAMD}\label{sec:compare}
We compare single HMAMD and single TOAMD wave functions for the central-type
correlation. The single HMAMD wave function with an imaginary shift $\bm{D}$ is
written as
\begin{equation}
\begin{aligned}
&\Psi_{\rm HMAMD}^{J\pi}(\bm{D})=\sum_{i<j}^A C_{ij} P^{J\pi}
{\cal A}
    \left\{
        \phi_i(\bm{r}_i,i\bm{D}) \phi_j(\bm{r}_j,-i\bm{D})
        \cdot \prod_{p\neq i,j}^{A-2}
        \phi_p(\bm{r}_p) 
    \right\}, \label{eq:hmamd-2}
\end{aligned}
\end{equation}
where $\cal A$ is an antisymmetrizer. $\phi_{i}$, $\phi_{j}$ and $\phi_{p}$ are
single-nucleon states, where $\phi_{i}$ and $\phi_{j}$ have imaginary shifts
$\pm i\bm{D}$ as shown in Eq.~(2) and $\phi_{p}$ is a single-nucleon state
expressed by Eq.~(\ref{eq:amd}). $P^{J\pi}$ denotes the angular momentum and parity
projections. 

When only the central correlation $F_S$ is included, for one Gaussian function term
of $F_S$ in Eq.~(\ref{eq:f-toamd-2}) with the range parameter $a$, we can express
the corresponding single TOAMD wave function as
\begin{align}
\Psi_{\rm TOAMD}
&= F_S\Psi_{\rm AMD}\\
&=\sum_{i<j}^A f_{ij} P^{J\pi} {\cal A}
    \left\{ 
        \prod_{p=1}^{A} \phi_p(\bm{r}_p)
    \right\},\label{eq:toamd-2}
\end{align}
where the pair function is $f_{ij}=e^{-a(\bm{r}_i-\bm{r}_j)^2}$.

We consider two assumptions in HMAMD below:
\begin{enumerate}
\item Superpose all the basis states having high-momentum pairs with an equal
weight, namely, $C_{ij}=1$ .
\item Integrate over the vector $\bm{D}$ corresponding to the momentum, with the 
Gaussian weight with the range $b$, namely, 
\begin{equation}
\Psi_{\rm HMAMD}^{J\pi} = \int d\bm{D} e^{-b\bm{D}^2} \Psi_{\rm
HMAMD}(\bm{D}) 
\end{equation}
\end{enumerate}
Under these assumptions, we consider the relation between the wave functions of
HMAMD and TOAMD. The integrated HMAMD wave function can be expressed as
\begin{equation}
\begin{aligned}
&\Psi_{\rm HMAMD}^{J\pi} = \int d\bm{D} e^{-b\bm{D}^2} \sum_{i<j}^A P^{J\pi} 
    {\cal A}
    \left\{ 
        \phi_i(\bm{r}_i,i\bm{D}) \phi_j(\bm{r}_j,-i\bm{D})
        \prod_{p\neq i,j}^{A-2} \phi_p(\bm{r}_p) 
    \right\}
\end{aligned}
\end{equation}
For the $s$-wave configuration, we limit the case of $\bm{R}=\bm{0}$ for all real
components of centroid $\bm{R}$ for $\phi_i$, $\phi_j$ and $\phi_p$. Hence from
Eq.~(\ref{single-state}), we have the relation
\begin{equation}
\begin{aligned}
&\phi_i(\bm{r}_i,i\bm{D}) \phi_j(\bm{r}_j,-i\bm{D}) = \phi_i(\bm{r}_i)
    \phi_j(\bm{r}_j) \times e^{2i\nu \bm{D}\cdot (\bm{r}_i-\bm{r}_j)}
    e^{2\nu\bm{D}^2},
\end{aligned}
\end{equation}
and substitute this relation into the single HMAMD wave function as
\begin{align}    
\Psi_{\rm HMAMD}^{J\pi}  
&= \sum_{i<j}^A P^{J\pi} {\cal A}
    \Bigg\{ 
    \int d\bm{D} 
        e^{-b\bm{D}^2}  
        e^{2i\nu \bm{D} \cdot (\bm{r}_i-\bm{r}_j)} 
        e^{2\nu\bm{D}^2}  \prod_{p}^{A} \phi_p(\bm{r}_p) 
    \Bigg\}\\
&= \sum_{i<j}^A P^{J\pi} {\cal A}\left\{e^{-a'(\bm{r}_i-\bm{r}_j)^2}
\cdot \prod_{p}^{A} \phi_p(\bm{r}_p) \right\}
\end{align}
where $a'=\frac{\nu^2}{b-2\nu}$. Considering $b$ is an adjustable parameter,
we may choose an appropriate $b$ to obtain $a'=a$ using a pair function $f_{ij}$, and
\begin{align}
\Psi_{\rm HMAMD}^{J\pi}  
&= \sum_{i<j}^A P^{J\pi} {\cal A}\left\{f_{ij} \prod_{p}^{A} \phi_p(\bm{r}_p)
\right\} \\
&= \sum_{i<j}^A P^{J\pi} \sum_p \epsilon(P) f_{P_i\, P_j}\cdot
\phi_1(\bm{r}_{P_1}) \phi_2(\bm{r}_{P_2})\cdots \phi_A(\bm{r}_{P_A}) \\
&= P^{J\pi} \sum_p \epsilon(P) \left( \sum_{i<j}^A f_{P_i\, P_j} \right) 
\phi_1(\bm{r}_{P_1}) \phi_2(\bm{r}_{P_2})\cdots \phi_A(\bm{r}_{P_A}).
\end{align}
Here, $\epsilon(P)$ is a sign for the permutation $P$ as
\begin{eqnarray}
  P~:~ \left(
\begin{array}{cccc}
1  &  2  & \cdots & A   \\
P_1& P_2 & \cdots & P_A 
\end{array}
\right).
\end{eqnarray}
From the symmetry of $f_{ij}$ we have the following relation
\begin{equation}
    \sum_{i<j}^A f_{P_i\, P_j}=\sum_{i<j}^A f_{ij}.
\end{equation}
Hence, this scalar factor can be factorized from the antisymmetrization and
$J^\pi$ projection operator as
\begin{align}
\Psi_{\rm HMAMD}^{J\pi} 
&=  \sum_{i<j}^A f_{ij} \cdot P^{J\pi} \cdot {\cal
A}\left\{\prod_{p}^{A} \phi_p(\bm{r}_p) \right\} \\
&=~F_S\Psi_{\rm AMD}\\
&=~\Psi_{\rm TOAMD}.
\end{align}
With this equation, we have shown the equivalence of the wave functions of the 
single HMAMD and single TOAMD for the $s$-wave configuration under the above two conditions.

\section*{References}
\def\JL#1#2#3#4{ {{\rm #1}} \textbf{#2}, #3 (#4)}  
\nc{\PR}[3]     {\JL{Phys. Rev.}{#1}{#2}{#3}}
\nc{\PRC}[3]    {\JL{Phys. Rev.~C}{#1}{#2}{#3}}
\nc{\RMP}[3]    {\JL{Rev. Mod. Phys.}{#1}{#2}{#3}}
\nc{\PRA}[3]    {\JL{Phys. Rev.~A}{#1}{#2}{#3}}
\nc{\PRL}[3]    {\JL{Phys. Rev. Lett.}{#1}{#2}{#3}}
\nc{\NP}[3]     {\JL{Nucl. Phys.}{#1}{#2}{#3}}
\nc{\NPA}[3]    {\JL{Nucl. Phys.}{A#1}{#2}{#3}}
\nc{\PL}[3]     {\JL{Phys. Lett.}{#1}{#2}{#3}}
\nc{\PLB}[3]    {\JL{Phys. Lett.~B}{#1}{#2}{#3}}
\nc{\PTP}[3]    {\JL{Prog. Theor. Phys.}{#1}{#2}{#3}}
\nc{\PTPS}[3]   {\JL{Prog. Theor. Phys. Suppl.}{#1}{#2}{#3}}
\nc{\PTEP}[3]   {\JL{Prog. Theor. Exp. Phys.}{#1}{#2}{#3}}
\nc{\PRep}[3]   {\JL{Phys. Rep.}{#1}{#2}{#3}}
\nc{\PPNP}[3]   {\JL{Prog.\ Part.\ Nucl.\ Phys.}{#1}{#2}{#3}}
\nc{\JPG}[3]     {\JL{J. of Phys. G}{#1}{#2}{#3}}
\nc{\andvol}[3] {{\it ibid.}\JL{}{#1}{#2}{#3}}

\end{document}